\documentclass{article}

\usepackage{PRIMEarxiv}

\usepackage[utf8]{inputenc} 
\usepackage[T1]{fontenc} 
\usepackage{hyperref} 
\usepackage{url} 
\usepackage{amsfonts} 
\usepackage{microtype} 
\usepackage{fancyhdr} 
\usepackage{graphicx} 
\usepackage[superscript, biblabel]{cite}
\usepackage{setspace}
\usepackage{lineno}
\usepackage{float}
\usepackage{xcolor}
\usepackage{appendix}

\title{Deep-learning-based clustering of oct images for biomarker discovery in age-related macular degeneration (pinnacle study report 4)
}

\begin{document}
\maketitle

\thispagestyle{empty}
\begin{singlespace}


\noindent \textbf{Authors:} 
Robbie Holland\footnote{BioMedIA, Imperial College London, London, United Kingdom} (robert.holland15@imperial.ac.uk), 
Rebecca Kaye \footnote{Clinical and Experimental Sciences, Faculty of Medicine, University of Southampton, Southampton, United Kingdom}, 
Ahmed M. Hagag\footnote{Institute of Ophthalmology, University College London, London, United Kingdom}$^{,}$\footnote{Moorfields National Institute for Health Research Clinical Research Facility, Moorfields Eye Hospital, London, United Kingdom}, 
Oliver Leingang\footnote{Laboratory for Ophthalmic Image Analysis, Medical University of Vienna, Vienna, Austria}, 
Thomas R. P. Taylor$^{2}$, 
Hrvoje Bogunovi\'c$^{5,}$\footnote{Christian Doppler Laboratory for Artificial Intelligence in Retina, Christian Doppler Forschungsgesellschaft, Vienna, Austria}, 
Ursula Schmidt-Erfurth$^5$, 
Hendrik P. N. Scholl \footnote{Institute of Molecular and Clinical Ophthalmology Basel, Basel, Switzerland} \footnote{Department of Ophthalmology, University of Basel, Basel, Switzerland}, 
Daniel Rueckert$^{1,}$\footnote{Institute for AI and Informatics in Medicine, Technical University Munich, Munich, Germany}, 
Andrew J. Lotery$^2$,
Sobha Sivaprasad$^{3,4}$ and 
Martin J. Menten$^{1,9}$








\noindent \textbf{Acknowledgments:} The PINNACLE study is funded by a Wellcome Trust Collaborative Award (ref. 210572/Z/18/Z). The authors would like to acknowledge Toby Prevost for his contribution to the acquisition of funding by the Wellcome Trust.

\noindent \textbf{List of acronyms:}
\begin{itemize}
\item AMD -- Age-related macular degeneration
\item OCT -- Optical coherence tomography
\item DLS -- Double-layer sign
\item RPE -- Retinal pigment epithelium
\item iRORA -- incomplete RPE and outer retinal atrophy
\item cRORA -- complete RPE and outer retinal atrophy
\item MNV -- Macular neovascularisation
\item SSL -- Self-supervised learning
\item SDD -- Subretinal drusenoid deposits
\item MAE -- Mean average error
\end{itemize}

\end{singlespace}

\clearpage
\thispagestyle{empty}
\section*{Abstract}

\textbf{Purpose:} We introduce a deep-learning-based biomarker proposal system for the purpose of accelerating biomarker discovery in age-related macular degeneration (AMD).

\textbf{Design:} Retrospective analysis of a large dataset of retinal optical coherence tomography (OCT) images.

\textbf{Participants:} A total of 3,456 adults aged between 51 and 102 years of whom OCT images were collected under the PINNACLE project.

\textbf{Methods:} Our system proposes candidates for novel AMD imaging biomarkers in OCT. It works by first training a neural network using self-supervised contrastive learning to discover, without any clinical annotations, features relating to both known and unknown AMD biomarkers present in 46,496 retinal OCT images. To interpret the learned biomarkers, we partition the images into 30 subsets, termed clusters, that contain similar features. We conduct two parallel 1.5-hour semi-structured interviews with two independent teams of retinal specialists to assign descriptions in clinical language to each cluster. Descriptions of clusters achieving consensus can potentially inform new biomarker candidates.

\textbf{Main Outcome Measures:} We checked if each cluster showed clear features comprehensible to retinal specialists, if they related to AMD and how many described established biomarkers used in grading systems as opposed to recently proposed or potentially new biomarkers. We also compared their prognostic value for late stage wet and dry AMD against an established clinical grading system and a demographic baseline model.

\textbf{Results:} Overall, both teams independently identified clearly distinct characteristics in 27 of 30 clusters, of which 23 were related to AMD. Seven were recognised as known biomarkers used in established grading systems and 16 depicted biomarker combinations or subtypes that are either not yet used in grading systems, were only recently proposed, or were unknown. Clusters separated {}{incomplete from complete retinal atrophy}, intraretinal from subretinal fluid and thick from thin choroids, and in simulation outperformed clinically-used grading systems in prognostic value.

\textbf{Conclusions:} Using self-supervised deep learning we were able to automatically propose AMD biomarkers going beyond the set used in clinically established grading systems. Without any clinical annotations, contrastive learning discovered subtle differences between fine-grained biomarkers. Ultimately, we envision that equipping clinicians with discovery-oriented deep-learning tools can accelerate discovery of novel prognostic biomarkers.

\clearpage
\setcounter{page}{1}
\section{Introduction}
\label{sec:introduction}
Age-related macular degeneration (AMD) is the leading cause of irreversible vision loss in the developed world and is projected to affect nearly 300 million people by 2040 \cite{wong2014global_s}. Clinical grading systems \cite{bird1995international,klein2014harmonizing,ferris2005simplified,ferris2013clinical} drive the screening, diagnosis and monitoring of AMD using a small set of known imaging biomarkers in retinal optical coherence tomography (OCT). The early and intermediate AMD stages are characterised by increasing sizes of drusen, which are lipidic subretinal deposits that often precede conversion to late AMD. It is currently impossible to reliably predict if and when a patient will progress to late AMD, which is classified as either macular neovascularisation (MNV), identified by macular fluid, or geographic atrophy, indicated by retinal and RPE thinning and hypertransmission of OCT signal in the choroid. The small number of known biomarkers, which permit only a coarse stratification of patients, is limiting the prognostic ability of current grading systems. There is a clear and unmet need for new imaging biomarkers that better describe patients’ physiology and provide improved risk stratification.

Identifying new biomarkers, or subtypes of known ones, is a very challenging task. New understanding of the pathophysiology of a specific disease can lead to the discovery of new biomarkers \cite{mcdermott2013challenges}. More commonly, biomarkers are proposed by clinicians making anecdotal observations in small cohorts. Using this ad-hoc approach clinicians have recently proposed {}{double-layer sign} (DLS) \cite{shi2019predictive}, vitreo-macular interface abnormalities \cite{robison2009vitreomacular} and intraretinal hyperreflective foci \cite{fragiotta2021significance} as biomarkers for AMD. Before entering clinical practice, candidate biomarkers must undergo extensive testing of their predictive value, reproducibility and utility \cite{abramson2015methods,kessler2015emerging}. Since drusen were proposed and integrated into clinical grading systems two decades ago, no new biomarkers for early-stage disease have been fully established \cite{age2001age}. The reliance on anecdotal observations, coupled with the onerous process of biomarker discovery, is a major constraint limiting the rate of biomarker discovery in AMD.

To expedite the initial stage of biomarker discovery there has been an increased push to harness data-driven approaches \cite{alyass2015big}. However, retinal specialists have not been able to translate the growing amount of retinal data into increased rates of biomarker discovery. Systematically searching for potentially new disease subtypes in large populations of patients is a tedious and resource-intensive task. Identifying new subtypes involves finding subgroups of images or patients that share some newly proposed characteristic. However, dividing a dataset of only 10,000 retinal images into as many as 30 subgroups would require making at least 300,000 simultaneous pairwise comparisons. The limited capacity in visual working memory \cite{cowan2001magical,miller1956magical} makes manual identification and exhaustive analysis of common patterns and biomarkers on this scale practically infeasible for human experts.

Deep learning has shown potential for automating visual tasks at scale in medical datasets \cite{schmidt2018artificial}. Traditional supervised approaches train networks to predict if any known biomarkers are present in the image. This both constrains networks to train only on the subset of labelled images, and limits their ability to discover new biomarkers outside the set known to the clinical annotators. The self-supervised learning (SSL) paradigm \cite{gidaris2018unsupervised,zhang2016colorful,oord2018representation,chen2020simple,he2020momentum,grill2020bootstrap} surpasses these limitations by training networks using unlabelled images. To train networks without supervision, researchers designed \textit{self}-supervised tasks where networks learn features through \textit{self}-discovery with minimal human input. Examples of these tasks include image reconstruction \cite{kingma2013auto} and contrastive predictive encoding \cite{oord2018representation}. After training, these learned features can be interpreted and potentially inform clinicians of new biomarkers that were present, but previously unseen, in the data.

\begin{figure}[t]
\centering
\includegraphics[width=0.99\textwidth]{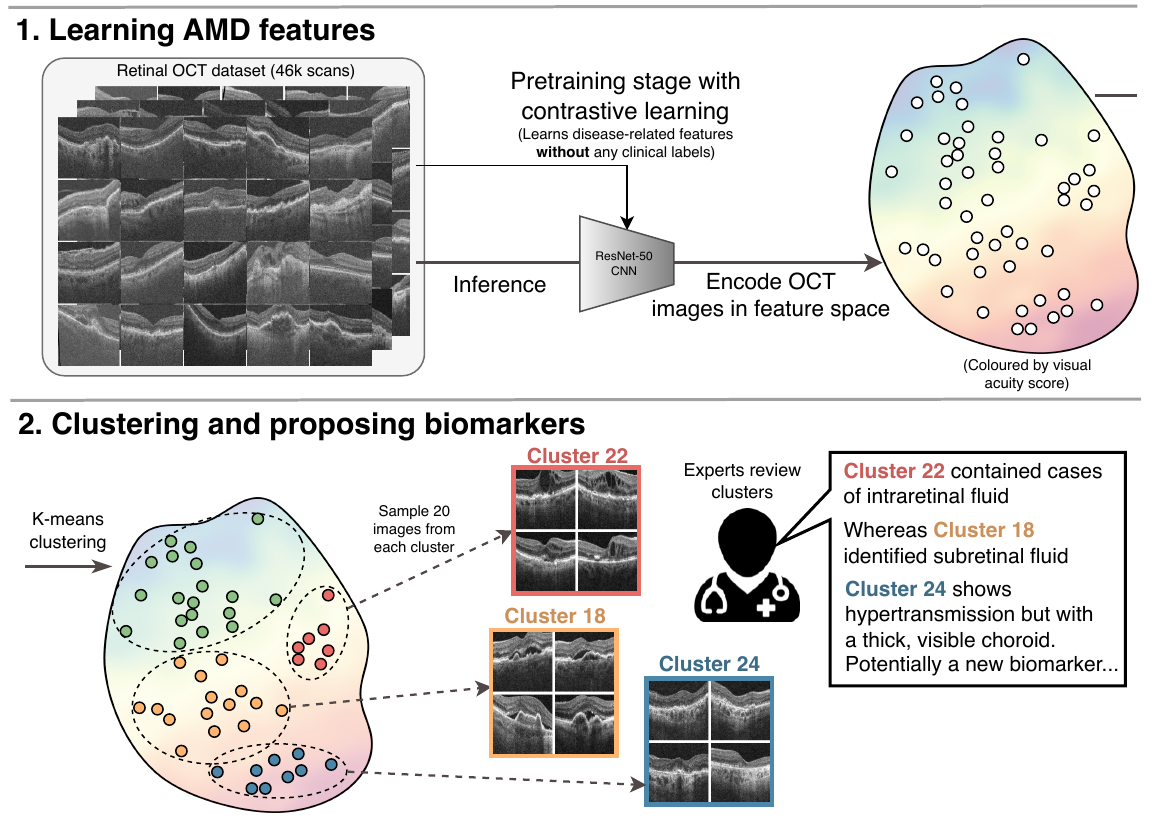}
\caption{In this study we design a biomarker proposal system based on contrastive learning.  After self-supervised pretraining, we cluster images that shared similar sets of features. Finally, two teams of retinal specialists independently identify the characteristic features of each cluster that potentially relate to new biomarkers.}
\label{fig:study_overview}
\end{figure}

Prior applications of SSL for biomarker discovery mistakenly incentivised networks to learn features that are uncorrelated with disease. Image reconstruction \cite{seebock2018unsupervised} and generative adversarial tasks \cite{schlegl2019f} prioritise learning features unrelated to imaging biomarkers, such as position and orientation of the retina in the image, and global image brightness and contrast. These features dominate the learning process and obfuscate subtler and more fine-grained indications of new biomarkers. Recent advances in SSL by contrastive methods address this issue by training networks to explicitly ignore these irrelevant features \cite{chen2020simple,he2020momentum,grill2020bootstrap}. As a result, they extract more robust and semantically relevant image features. Despite their potential for discovering new imaging features related to disease, contrastive methods have received little to no attention for biomarker discovery.

In this work we develop the first contrastive-learning-based proposal system for AMD imaging biomarkers. By pretraining a network with contrastive learning, we were able to automatically extract self-supervised features from 46,496 OCT images of 6,236 eyes from 3,456 patients without any clinical supervision. We then automatically identified 30 distinct groups of images, termed clusters, that share similar features. Two teams of retinal specialists independently reviewed the clusters, assigning clinical interpretations to each. Finally, we report clusters characterised by unique sets of features relating to known, suspected and unknown AMD imaging biomarkers.

\section{Methods}
The workflow used in this study is illustrated in Figure \ref{fig:study_overview}. We begin by introducing the OCT dataset and the set of known biomarkers in section \ref{sec:dataset}. In the first stage of the automatic discovery process we learn imaging features from this dataset using self-supervised contrastive learning, detailed in section \ref{sec:contrastive}. In the second stage we automatically identifying subgroups, termed clusters, of images with shared features. Treating those shared features as possible biomarker candidates, we describe our strategy to interpret them in section \ref{sec:interpreting}. Finally, we outline our full procedure for evaluating our method in section \ref{sec:evaluation}.

\subsection{OCT dataset}
\label{sec:dataset}
Experiments were conducted using a dataset of 46,496 OCT images of of 6,236 eyes from 3,456 patients acquired over an eight-year period at the Southampton Eye Unit and collected by the PINNACLE consortium \cite{sutton2022developing}. {}{The PINNACLE study (ClinicalTrials.gov NCT04269304) protocol was approved in the United Kingdom by the East Midlands–Leicester Central Research Ethics Committee (ref. 19/EM/0163) and further by the institutional review boards of all involved institutions. It adheres to the principles of Good Clinical Practice and is in accordance with the Declaration of Helsinki.} All images were acquired using Topcon 3D OCT scanners (Topcon Corporation, Tokyo, Japan). From each we extracted a 2D mediolateral slice centered on the fovea, adjusted to a size of $208\times256$ pixels, and set the pixel size to $7.0\times23.4\mu m^2$, representing half the median resolution. Visual acuity LogMAR scores, which measured the patient's functional quality of vision, were measured at 34,093 visits. These were converted to Letter Scores, indicating how many letters the patient can read on a chart (from 5-95). 

\noindent \textbf{Current grading system annotations}: A subset of 7,730 images from 1,031 eyes {}{from 1,001 patients} were labelled by the retinal specialists using the established AMD grading protocols that derive from known imaging biomarkers. Early/intermediate AMD was characterised by drusen at least $63 \mu m$ in diameter. We also recorded MNV, cRORA (atrophy of width $\geq 250\mu m$ and $< 1000\mu m$), larger cRORA (atrophy of width $\geq 1000\mu m$) \cite{sadda2018consensus} and healthy cases with no visible pathological biomarkers.

\subsection{Self-supervised contrastive learning}
\label{sec:contrastive}
\begin{figure}[t]
\centering
\includegraphics[width=\textwidth]{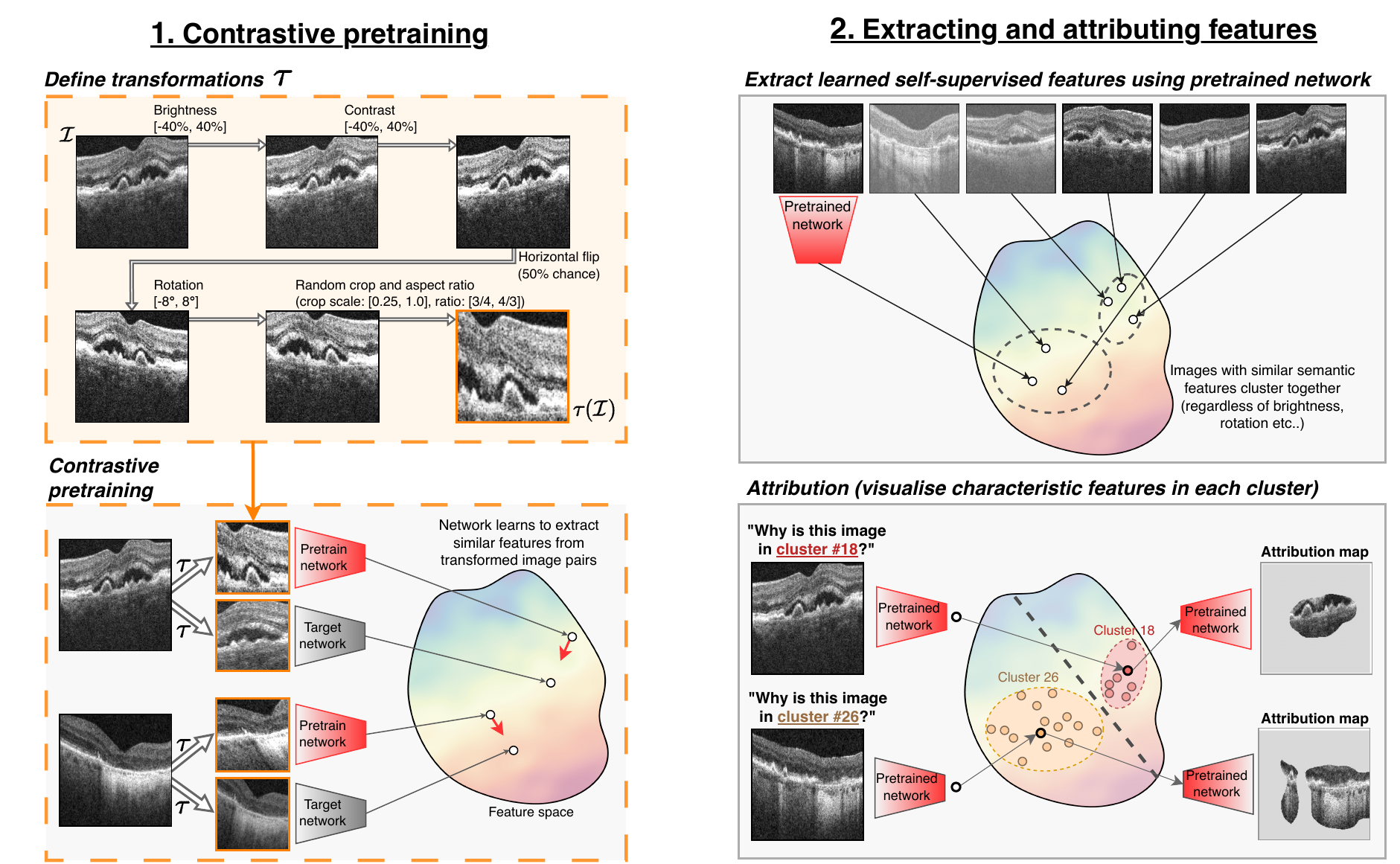}
\caption{The fully automated backbone of our biomarker proposal system consists of two stages. Firstly, self-supervised contrastive learning trains models to identify biomarkers and other image features without any clinical annotations. To do so, it trains networks to ignore a specified set of known invariant image features, defined by the set of contrastive transformations, amplifying the signal of any biomarker related features. Secondly, we extract self-supervised image features and cluster images with similar features. After this we compute attribution maps that highlight the cluster-specific features in each image to assist interpretation by retinal specialists.}
\label{fig:contrastive}
\end{figure}

We used contrastive learning to train a self-supervised feature extractor $f$ which will form the basis of our biomarker proposal system. Contrastive learning works by creating two augmented versions of each image using a pre-defined set of transformations (see Figure \ref{fig:contrastive}). By training $f$ to extract a similar set of features from each augmented version, $f$ learns to discern imaging features that are invariant to the transformations. By intentionally using transformations that do not alter the presentation of structural imaging biomarkers, we aimed to increase both the amount and subtlety of learned features that relate to potentially new biomarkers. To this end, we varied global image brightness and contrast, image rotation and aspect ratio, horizontal symmetry and a randomly sized and located crop. The latter requires $f$ to relate features that co-occur across the OCT images, such as hypertransmission (in the choroidal region) and retinal thinning. The exact parameters used for these transformations are tailored for retinal OCT by Holland et al. \cite{DBLP:journals/corr/abs-2208-02529}.

We choose the contrastive loss from Bootstrap Your Own Latent \cite{DBLP:journals/corr/abs-2208-02529} which trains $f$ to minimise the distance in feature space between the two transformed image versions. We parameterise $f$ by a ResNet50 (4x) network \cite{he2016deep}, and train for 120,000 steps using the Adam optimiser with a momentum of 0.9, weight decay of $1.5 \cdot 10^{-6}$ and a learning rate of $5 \cdot 10^{-4}$. Once training was complete, we first removed the final linear layer of $f$ before extracting 2048 unsupervised features from each labelled image. This so-called feature space, or vector database, compactly encoded each image's most salient features for further analysis.

\subsection{Clustering and feature attribution}
We employed a cluster-based approach to identify subgroups of self-supervised features learned by $f$. These clusters are later reviewed by clinicians, who give clinical interpretations to their characteristic sets of features. To perform the clustering, we used k-means to assign each image to one of 30 distinct clusters. We then used three annotation strategies to enhance the interpretability of the clusters prior to review. Firstly, we reordered clusters by their median visual acuity, so that higher cluster numbers indicated a more degraded quality of functional vision. We referred to our clusters using the schema C1 (best vision) up to C30 (worst vision). Secondly, we correlated each cluster with a disease stage from the current grading system. To do this we found the probability that any given image from each cluster was annotated with each of the known biomarkers. Thirdly, we highlighted the features in each image that are most representative of its assigned cluster (see Figure \ref{fig:contrastive}). To this end, we fit a single linear layer that mapped from the feature space to cluster assignments. Then, to highlight the cluster-specific image features we generated an \textit{attribution map} by applying GradCAM \cite{selvaraju2017grad} between the image and the cluster label using the final two convolutional layers of $f$.

\subsection{Interpreting clusters with retinal specialists}
\label{sec:interpreting}
Until this stage, the entire automated discovery process had not used any clinical annotations or manual input. We then aimed to understand and interpret the features that defined and distinguished each cluster. To this end, we recruited two independent teams of retinal specialists, each consisting of a senior and a junior ophthalmologist, and conducted two parallel 1.5 hour semi-structured interviews. For each cluster, we first showed the specialists 10 randomly drawn images and their associated attribution maps, ensuring all images derived from different patients. By comparing multiple images from the same cluster, the retinal specialists could visually extract, distil, and articulate any image features that represented the majority. While interpreting the images, the specialists were also permitted to inspect other clusters that correlated with the same disease grading. This allowed them to better identify image features that are not currently differentiated by established grading systems but were, by definition, distinguishable by their self-supervised features. This could enable the detection of potentially new biomarker subtypes that are currently conflated in each indiscriminate disease stage.

After reviewing the 10 images, each team suggested up to three common features using clinical terms, such as large drusen or poor image quality, for each cluster and ranked them according to their prevalence among the images. If there is no identifiable feature describing the majority of images, the cluster was labelled `heterogeneous'. To validate these descriptions, a further set of 10 unseen images and attribution maps were subsequently revealed. The suggested features were only taken as the team's proposed descriptions if they also described the majority of the unseen images. Finally, if there is semantic consensus between the two independent teams' descriptions, we reported this as the overall description of the cluster. Cluster descriptions may reveal potentially new biomarkers, providing the basis for proposing novel biomarker candidates.

\subsection{Evaluation}
\label{sec:evaluation}
Our main evaluation assessed the ability of our system to propose novel biomarkers. To assess this, we determined how often both expert teams identified the same features in each of the 30 clusters. Next, we recorded the number of clusters where the identifiable features related to biomarkers, as a measure of the ability of contrastive learning to ignore disease-unrelated image features. Finally, we observed how many of those biomarkers were already included in existing grading systems, as opposed to those that have been proposed by clinical researchers, but not yet included in grading systems, or were previously unknown. These statistics are our predominant measure for the effectiveness of our biomarker proposal system.

In addition to being discernible in images, biomarkers should provide some degree of risk stratification and prognostic value for late AMD. We compared the predictive value of our clusters to the known biomarkers, and a demographic baseline that uses only the patient's age and sex. To make predictions using our clusters, each image was represented by a vector of size 30 that encodes its similarity to each cluster. This vector was then used by a Lasso linear regression model to estimate the risk. We quantify each system's prognostic capability by calculating the Mean Absolute Error (MAE) in years to forecast the time until a patient's conversion to MNV or cRORA. In addition, we predicted the current visual acuity. Finally, we included a fully supervised learning baseline by fitting a linear support vector regression (SVR) model directly to the feature space to demonstrate the performance gap between our interpretable cluster-based approach and black box models. Each experiment used 10-fold cross validation on random 80/20 partitions, while ensuring a patient-wise split. We repeated the entire experiment, from clustering to regression, using 7 random seeds and report means and standard deviations.

\section{Results}

\subsection{Quality of clusters}

\begin{figure}[]
\centering
\includegraphics[height=0.93\textheight]{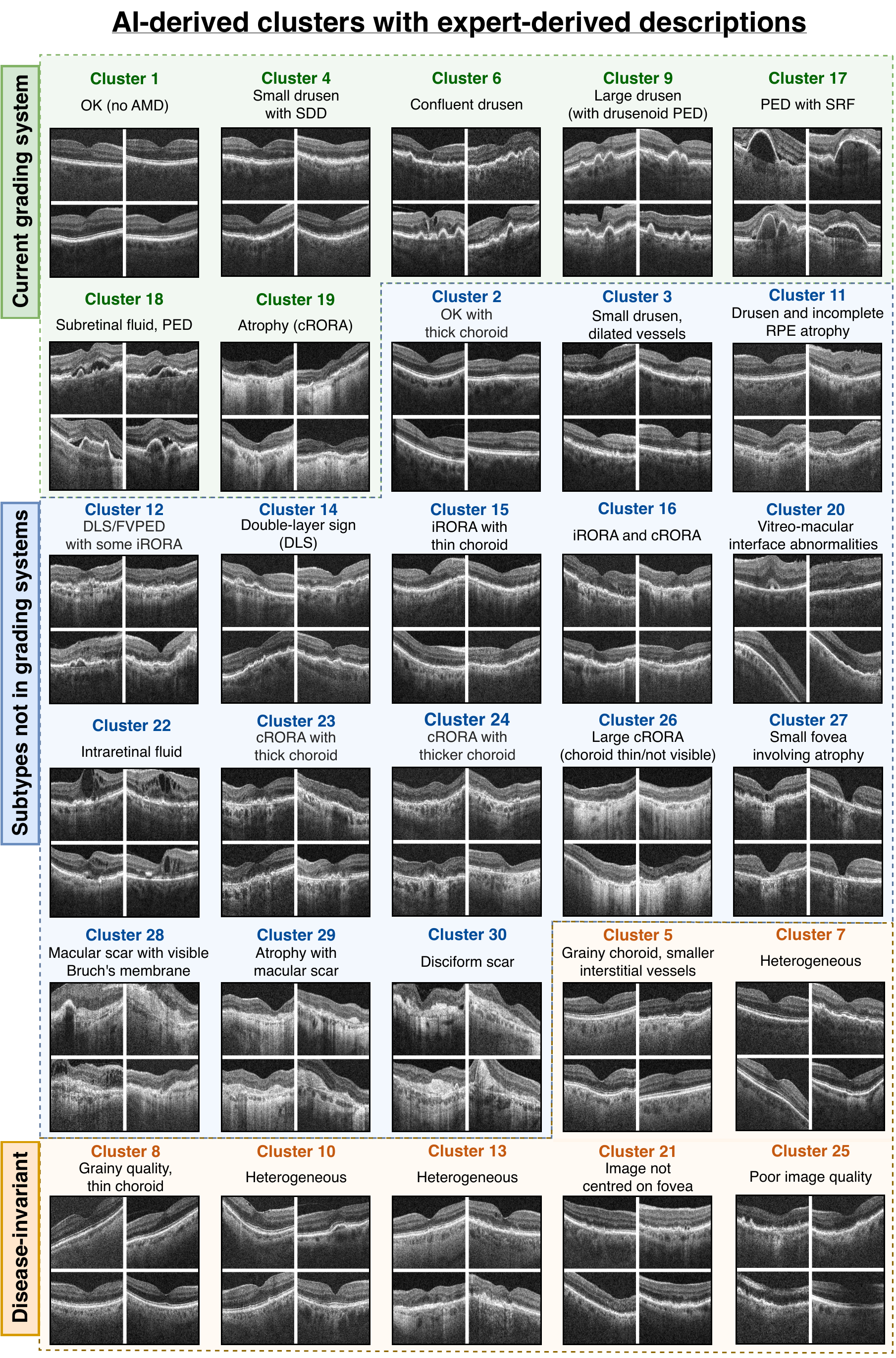}
\caption{Each cluster with its description derived independently by two teams of retinal specialists. In each we show four representative images from different patients. Our proposal system identified these clusters without any human supervision or prior knowledge of known biomarkers through a process of self-discovery. Out of 30 clusters, 23 were related to AMD of which 16 made subtle distinctions between fine-grained biomarkers that were either unknown to retinal specialists or not included in existing clinical grading systems.}
\label{fig:clusters}
\end{figure}

Both teams of retinal specialists independently reported semantically identical descriptions in 27 out of 30 clusters (see Figure \ref{fig:clusters}), and remarked on the consistency of image features observed within each. The two teams disagreed, or found no clearly identifiable features, in only three clusters (C7, C10, C13) deemed heterogeneous. Notably, 23 clusters were found to capture different biomarkers related to AMD. We have approximately divided these into a group of 7 clusters representing biomarkers already in current grading systems, and a group of 16 that exhibit potentially new subtypes. In a minority of clusters (C5, C8, C21, C25) the common features were instead related to image or scan quality. Importantly, there were no clusters for image brightness, rotation and position of the retina in the image, or any feature explicitly included in the set of contrastive transformations. The high proportion of clusters depicting identifiable imaging biomarkers confirms our hypothesis that contrastive learning is able to extract high-level semantic features and ignore many confounding image features.

\begin{figure}[ht!]
\centering
\includegraphics[width=0.95\textwidth]{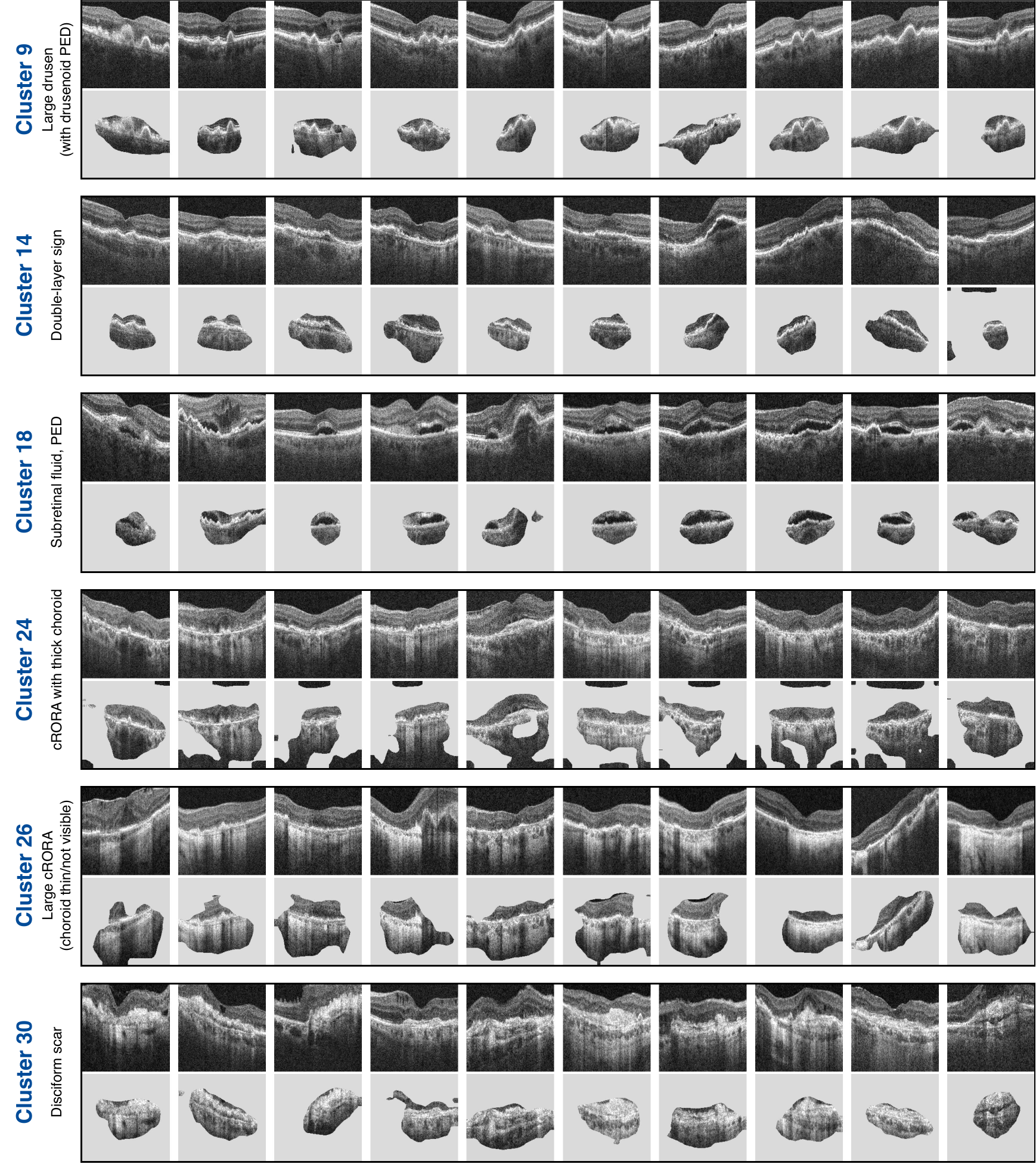}
\caption{Images from ten randomly drawn patients from six clusters, as shown to retinal specialists during the cluster interpretation interviews. The clusters were were largely homogeneous and had identifiable features that described the majority of the images (written to the left). This was reinforced by cluster-specific attribution maps (below each image) that indicating that a consistent set of self-supervised features define each cluster.}
\label{fig:all_samples}
\end{figure}

{}{Each cluster contained at least 120 images, originating from at least than 46 unique patients (see Table \ref{tab:cluster_statistics}). This indicated that the clusters captured biomarkers that arose independently among multiple patients with AMD, rather than features that were specific to only a few outlier patients.} This is also shown in Figure \ref{fig:all_samples} where, in most clusters, images from ten randomly drawn patients presented consistent features related to AMD that were distinct from other clusters. This was reinforced by GradCAM attribution maps which, for example, highlighted drusen in C9, hypertransmission in C26 and subretinal hyperreflective material (disciform scars) in C30. 

\subsection{Clusters describe known biomarkers}
Both teams of retinal specialists found clusters that distinguished healthy retinas (C1), small drusen with subretinal drusen deposits (SDD) (C4), confluent drusen (some post anti-vegf treatment) (C6) and large drusen (C9). Drusen size is an established predictor for risk of progression to late AMD \cite{abdelfattah2016drusen, schlanitz2017drusen}. Moreover, they identified clusters described as subretinal fluid (C18) and pigment epithelial detachment (C17), or PED, which are both indicative of MNV. Finally, images in cluster C19 clearly exhibit atrophy or cRORA. These clusters represent the known biomarkers used in clinical grading systems that currently map the progression of early AMD and its wet and dry variants in late AMD.

\subsection{Clusters contain biomarkers not yet in established grading systems}
\begin{figure}[ht]
\centering
\includegraphics[width=\textwidth]{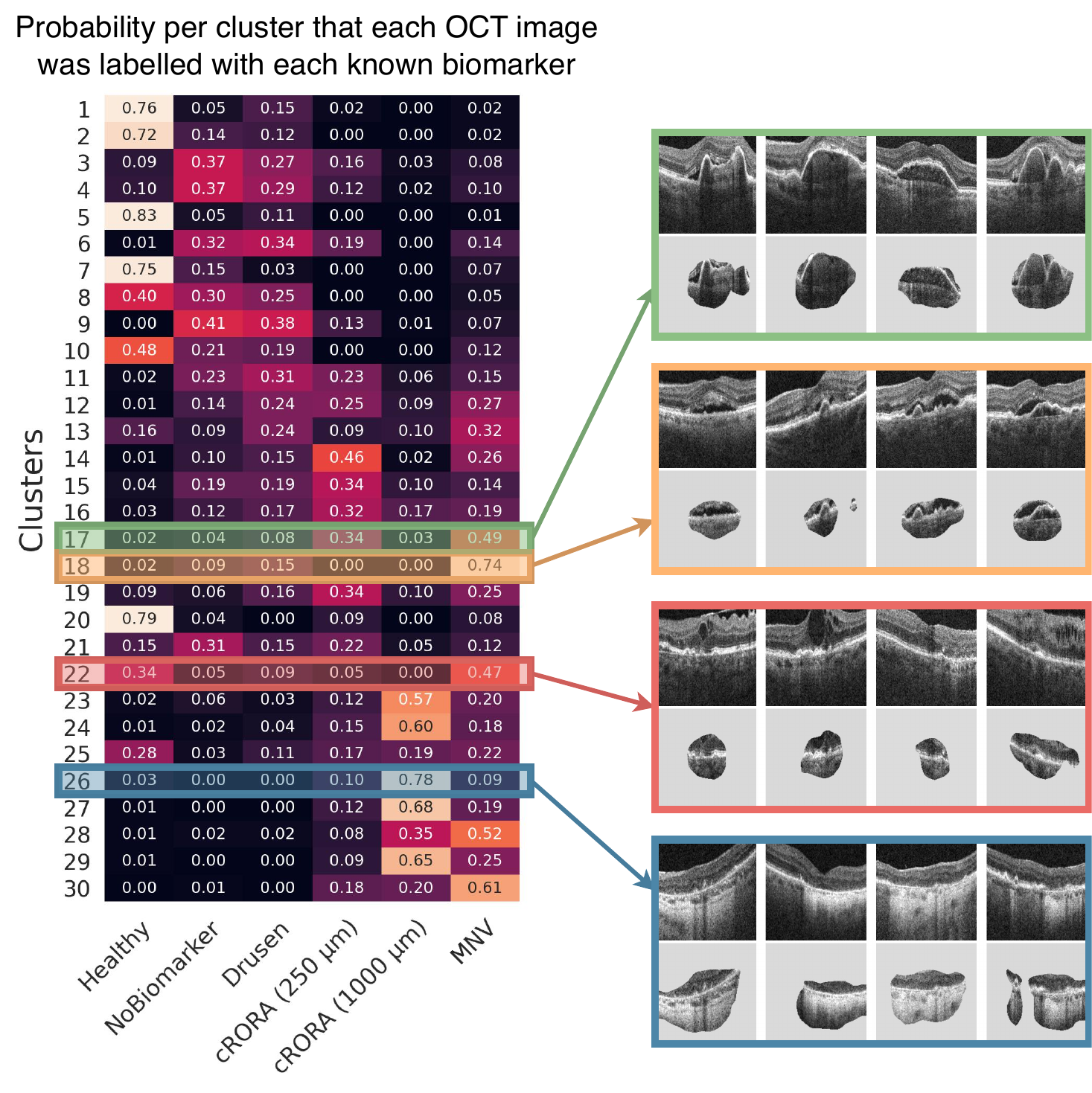}
\caption{Clusters were correlated with known biomarker annotations and disease stages using conditional probability. By comparing clusters that are indistinguishable to current grading systems but are, by definition, distinguishable by their self-supervised features we hope to identify new biomarker subtypes that are currently conflated in each indiscriminate disease stage.}
\label{fig:conditional_probability}
\end{figure}

The majority of the clusters distinguished cases that would fall under the same category in established grading systems. As shown in Figure \ref{fig:conditional_probability}, our clusters create a more fine-grained decomposition of the more broadly defined known biomarkers. Moreover, correlating clusters with known disease stages also highlighted a subgroup, C20 with vitreo-macular interface abnormalities, that had been mislabelled as healthy. Our clusters differentiated intraretinal fluid (C22) from subretinal fluid (C18). These are typically conflated under neovascular AMD \cite{mitchell2018age}, despite the known association between intraretinal fluid and degraded visual acuity \cite{jaffe2013macular,sharma2016macular,kaiser2021retinal}, which was reflected in our own measurements where average visual acuity for C22, $58.7$ letters (95\% CI $[55.9, 61.4]$), was worse than in C18, $68.2$ letters (95\% CI $[66.7, 69.8]$) (see Figure \ref{fig:valogmar}).

Other clusters described biomarkers that have only been tentatively linked to the progression of AMD such as double-layer sign (C14), which is associated with subclinical type 1 macular neovascularisation \cite{shi2019predictive}, and vitreo-macular interface abnormalities (C20) including epiretinal membrane, which has been shown to increase the growth rate of geographic atrophy \cite{grunwald2015growth}. Moreover, our clusters reflected the distinction in the severity of atrophy between {}{incomplete (C12, C15) and complete (C24, C26) retinal pigment epithelial and outer retinal atrophy (iRORA and cRORA)} that was only recently proposed in 2018 \cite{sadda2018consensus}. Clusters also differentiated incomplete RPE atrophy (C11) from iRORA and cRORA. One cluster was characterised by small fovea involving atrophy and macular holes (C27) and had the worst visual acuity of any cluster not depicting scarring. Fovea-centered atrophy has been shown to have the most deleterious effects on visual acuity \cite{bagheri2017foveal}. The subtlety of these distinctions supports our hypothesis that features learned by contrastive learning can extract fine-grained biomarkers.

\subsection{Clusters describe potentially new biomarkers}
Some clusters were differentiated by the presence of a thick, visible choroid. For example, in contrast to C1, C2 contained healthy retinas but with thick, permeable choroid. Similarly, C16 showed iRORA with a thick choroid in contrast to C15 exhibiting iRORA with a thin choroid. Both teams of retinal specialists were most interested in C24 which showed cRORA but with a thick, visible choroid. This was unexpected, as usually atrophy is accompanied with a thin choroid as in C19 and C26. Poor choroidal perfusion of the macula is implicated in the pathogenesis of AMD \cite{mori2005decreased}, which may indicate that patients in clusters with thinner choroids are at higher risk. Another novel combination includes DLS and fibrovascular PED (C12). These findings evidence the capability of our system to propose biomarker subtypes that had not yet been considered.

\subsection{Risk stratification with clusters}
\begin{figure}[htbp]
\centering
\includegraphics[width=0.9\textwidth]{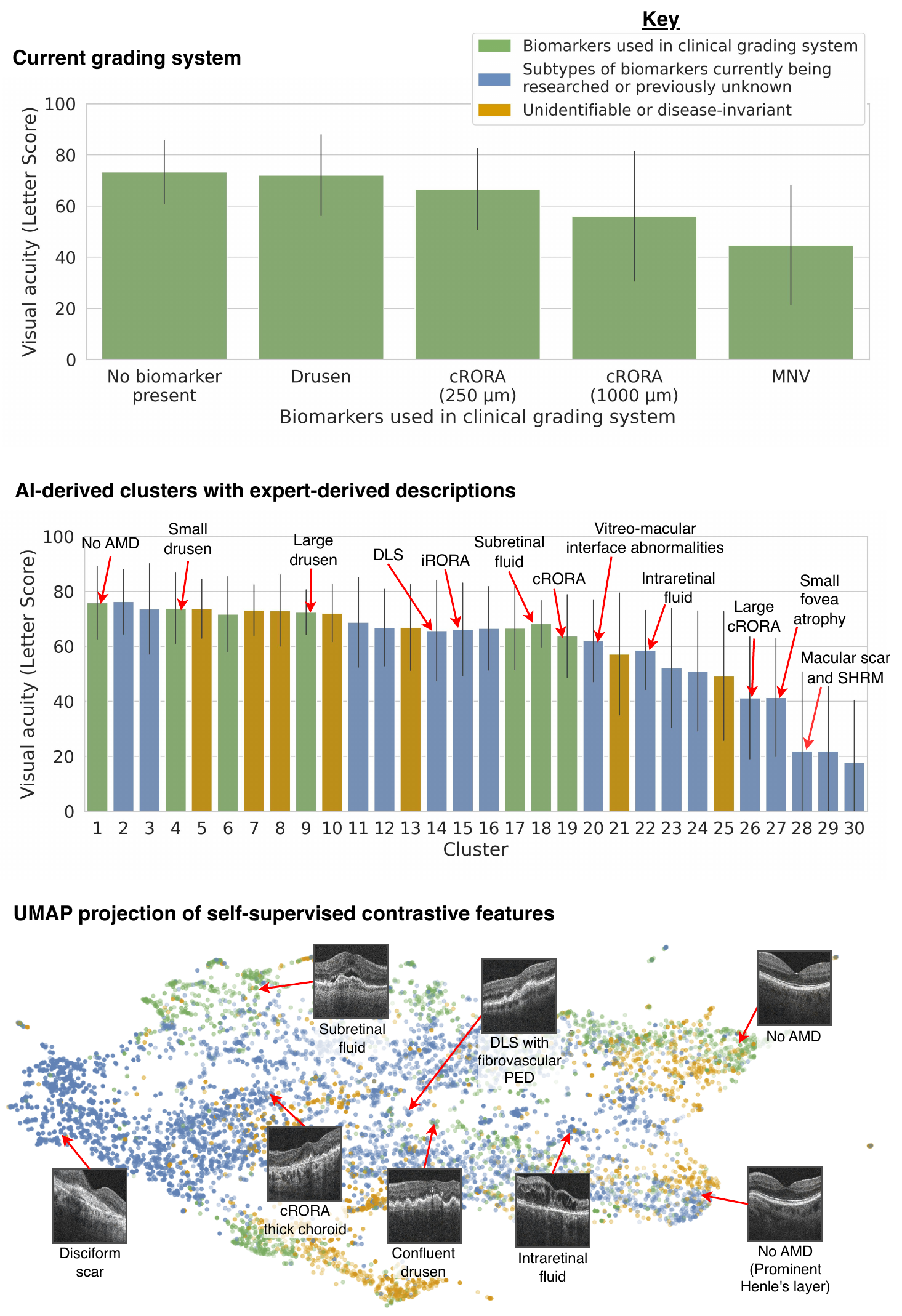}
\caption{Comparing the stratification of visual acuity (in Letter Score) by each stage of the clinical grading system cluster (top) to our clusters ({}{middle}). Each bar represents the average visual acuity with error bars for standard deviation. Our interpretable clusters provide a significantly improved stratification of degradation to visual acuity compared to the set of known biomarkers. {}{We also use a UMAP projection (bottom) to depict where retinal images reside in the self-supervised contrastive feature space used to create the clusters.}}
\label{fig:valogmar}
\end{figure}

By simulating a grading system using cluster membership, we found that the fine-grained biomarker-related distinctions made by our clusters were more prognostic than the broad disease stages in the clinical grading system. This was evidenced by reduced MAE in predicting time until conversion to late wet AMD and late dry AMD (shown in Table \ref{tab:quant}). Notably, the current grading system (defined in section \ref{sec:dataset}) did not stratify patient risk much better than the patient's age and sex. The fully supervised black-box model performed the best, but this approach makes uninterpretable predictions. We found our clusters were especially effective at predicting current visual acuity (11.5 vs. 18.4 MAE). This performance is enabled by the use of fovea-centred images that effectively capture the macula. Moreover, we find clusters provide more granular stratification of the degradation in visual acuity (see Figure \ref{fig:valogmar}). These results reinforce that our clusters are related to disease progression.

\begin{table}[t]
\centering
\caption{Clusters outperformed the established clinical grading systems for AMD in predicting future conversion to disease, shown by reduced mean average error (MAE) in years for Late AMD, MNV and cRORA. We also find clusters {}{substantially} reduced MAE in Letter Score for predicting visual acuity. Fully supervised machine learning outperforms our clusters, but is uninterpretable and cannot form better grading systems.}
\label{tab:quant}
\begin{tabular}{rcccc}
\multicolumn{1}{c|}{System}      & \multicolumn{1}{c|}{\begin{tabular}[c]{@{}c@{}}Time to\\ Late AMD \\ (MAE years) $\downarrow$\end{tabular}} & \multicolumn{1}{c|}{\begin{tabular}[c]{@{}c@{}}Time to\\ MNV \\ (MAE years) $\downarrow$\end{tabular}} & \multicolumn{1}{c|}{\begin{tabular}[c]{@{}c@{}}Time to\\ cRORA \\ (MAE years) $\downarrow$\end{tabular}} & \begin{tabular}[c]{@{}c@{}}Current\\ visual acuity \\ (MAE Letters) $\downarrow$\end{tabular} \\ \hline
\multicolumn{1}{r|}{Demographic}            & \multicolumn{1}{c|}{0.76$\pm$0.01} & \multicolumn{1}{c|}{0.82$\pm$0.01}            & \multicolumn{1}{c|}{0.70$\pm$0.03}     & 19.1$\pm$0.35     \\
\multicolumn{1}{r|}{Current grading system} & \multicolumn{1}{c|}{0.76$\pm$0.01} & \multicolumn{1}{c|}{0.82$\pm$0.01}            & \multicolumn{1}{c|}{0.69$\pm$0.04}     & 18.4$\pm$0.40     \\
\multicolumn{1}{r|}{Clusters}               & \multicolumn{1}{c|}{0.75$\pm$0.01}& \multicolumn{1}{c|}{0.78$\pm$0.02}            & \multicolumn{1}{c|}{0.63$\pm$0.05} & 11.5 $\pm$0.25      \\
\multicolumn{1}{r|}{Fully supervised}       & \multicolumn{1}{c|}{0.71$\pm$0.02}& \multicolumn{1}{c|}{0.73$\pm$0.01}            & \multicolumn{1}{c|}{0.61$\pm$0.03}     & 10.0 $\pm$0.20    
\end{tabular}
\end{table}

\section{Discussion}
In this study we introduced a contrastive-learning based approach to biomarker discovery in AMD. We hypothesised that contrastive learning could reveal new biomarkers by focusing deep-learning models on potentially new biomarkers. Through a process of automated self-discovery our system rediscovered known biomarkers, and proposed candidates for potentially new ones. Without using any clinical annotations we automatically grouped a dataset of over 46,000 retinal OCT images into 30 distinct clusters. In only two 1.5 hour interviews, two teams of retinal specialists empowered with our tool built a granular taxonomy of the AMD progression in a dataset. The teams independently verified that clusters modelled fine-grained distinctions between various subtypes of AMD. They remarked both on the consistency in shared features between images of the same cluster, and on the subtlety of the different biomarkers captured by different clusters. In addition to finding tentative biomarkers that are not yet in current grading systems, retinal specialists were surprised to find clusters depicting nuanced and granular subtypes of biomarkers that had not yet been identified in clinical research. These subtypes included new combinations such as thick choroid with hypertransmission, double-layer sign seen with a fibrovascular PED, and small fovea involving atrophy. Moreover, in quantitative benchmarks our clusters provided greater risk stratification and prognostic value than the current clinical grading system. Our results strongly suggest that biomarker proposal powered by contrastive learning can accelerate the initial idea-generation stage of biomarker discovery. 

Our approach offers a data-driven and scalable alternative to traditionally anecdotal biomarker proposals. It provides a fast, effective and inexpensive preliminary review of any medical dataset, which can subsequently be refined using fewer resources. It can also enable the analysis of the compound effect on progression risk posed by enumerable biomarker combinations and subtypes. {}{Our method, from pretraining the feature extractor, to clustering and finally clinical interpretation, is not specific to any disease, anatomical region or scanning instrument. However, if a user has images acquired using a different domain, such as a change in OCT instrument or pathology of interest, they should first retrain the feature extractor on data from their specific domain before identifying new biomarker proposals.} By integrating our tool into existing biomarker discovery pipelines, advances in deep learning can improve patient care without deploying a single black-box neural network to the clinic.

Although our clusters were found to be largely homogeneous, there were instances where not every image within a cluster shared the majority feature. We emphasise that clusters should be treated as biomarker proposals requiring further refinement by clinicians, rather than as definitive biomarkers. They are intended as preliminary suggestions that provide a starting point for further investigation. Consequently, our quantitative study only indicates that a more nuanced set of biomarkers may enhance risk stratification for AMD. This also complicates measuring definitive quantities of our clusters, such as the minimum width of atrophy (such as $cRORA > 1000\mu m$). However, all the aforementioned issues can be addressed by manually refining each cluster.

Moreover, clusters are not guaranteed to be characterised by biomarker-related features. Of the 30 clusters, 4 related to invariant image features, such as scan location, that are difficult to model in the set of contrastive transformations and may require a more sophisticated technique to explicitly remove from the set of self-supervised features. For now, these clusters can easily be removed during the review stage and could even be filtered from the dataset as a preprocessing step (for example, removing C21 would omit acquisitions not centered on the fovea). However, in another 3 clusters the retinal specialists were unable to identify any majority feature. This could either indicate the existence of catch-all heterogeneous clusters of miscellaneous images, or that both teams of retinal specialists missed potentially new biomarker subtypes. As two of these clusters had no obvious signs of AMD, it may also indicate that identifying early-stage biomarkers, which produce more subtle structural deformities, can be more challenging than detecting more pronounced late-stage ones. Improving the chance of detecting new early-stage biomarkers may require filtering late-stage patients from the process. A more definitive limitation of our system is that by using 2D fovea-centered B-scans, we potentially miss pathologies in other regions of the retina. {}{This potentially missed pathology may explain some of the error in predicting visual acuity, which was 10 letters using fully supervised learning, equating to two lines in a LogMAR chart.} To incorporate biomarkers outside the macula and enhance the prognostic value of our clusters, we plan to repeat this study using the full 3D OCT volume.

\noindent \paragraph{Conclusion} In this work we introduced a deep-learning-based biomarker proposal system powered by self-supervised contrastive learning, capable of identifying biomarker candidates in large medical image datasets with minimal human input. Our system, without any prior knowledge of AMD, rediscovered all the known biomarkers. Crucially, it also proposed biomarker candidates that had either only been raised in a small number of papers, or had not been yet identified by retinal specialists to the best of our knowledge. Ultimately, we envision deep-learning-based self-discovery systems can improve patient outcomes by accelerating the rate of biomarker research into this poorly understood disease.

\newpage

\bibliographystyle{ama}
\bibliography{manuscript}








\newpage
\clearpage
\begin{appendix}
\renewcommand{\thetable}{A.\arabic{table}}
\setcounter{table}{0}
\section*{Supplementary material}
\begin{table}[htp!]
\centering
\caption{{}{The number of images contained in each cluster, in addition to the number of unique patients the images originate from. In most clusters, the average patient contributed no more than 3 images. Clusters C29 and C30, which had higher ratios, related to scarring and largely document patients being monitored post-treatment.}}
\label{tab:cluster_statistics}
\begin{tabular}{rccccccccccccccc}
\hline
\multicolumn{1}{|r|}{\textbf{Cluster}}    & \multicolumn{1}{c|}{\textbf{1}}  & \multicolumn{1}{c|}{\textbf{2}}  & \multicolumn{1}{c|}{\textbf{3}}  & \multicolumn{1}{c|}{\textbf{4}}  & \multicolumn{1}{c|}{\textbf{5}}  & \multicolumn{1}{c|}{\textbf{6}}  & \multicolumn{1}{c|}{\textbf{7}}  & \multicolumn{1}{c|}{\textbf{8}}  & \multicolumn{1}{c|}{\textbf{9}}  & \multicolumn{1}{c|}{\textbf{10}} & \multicolumn{1}{c|}{\textbf{11}} & \multicolumn{1}{c|}{\textbf{12}} & \multicolumn{1}{c|}{\textbf{13}} & \multicolumn{1}{c|}{\textbf{14}} & \multicolumn{1}{c|}{\textbf{15}} \\ \hline
\multicolumn{1}{|r|}{\textbf{\#Images}}   & \multicolumn{1}{c|}{310}         & \multicolumn{1}{c|}{262}         & \multicolumn{1}{c|}{400}         & \multicolumn{1}{c|}{370}         & \multicolumn{1}{c|}{234}         & \multicolumn{1}{c|}{249}         & \multicolumn{1}{c|}{166}         & \multicolumn{1}{c|}{166}         & \multicolumn{1}{c|}{198}         & \multicolumn{1}{c|}{197}         & \multicolumn{1}{c|}{395}         & \multicolumn{1}{c|}{268}         & \multicolumn{1}{c|}{205}         & \multicolumn{1}{c|}{233}         & \multicolumn{1}{c|}{435}         \\ \hline
\multicolumn{1}{|r|}{\textbf{\#Patients}} & \multicolumn{1}{c|}{196}         & \multicolumn{1}{c|}{173}         & \multicolumn{1}{c|}{166}         & \multicolumn{1}{c|}{174}         & \multicolumn{1}{c|}{156}         & \multicolumn{1}{c|}{102}         & \multicolumn{1}{c|}{114}         & \multicolumn{1}{c|}{103}         & \multicolumn{1}{c|}{69}          & \multicolumn{1}{c|}{121}         & \multicolumn{1}{c|}{170}         & \multicolumn{1}{c|}{110}         & \multicolumn{1}{c|}{126}         & \multicolumn{1}{c|}{69}          & \multicolumn{1}{c|}{174}         \\ \hline
\multicolumn{1}{|r|}{\textbf{Ratio}}      & \multicolumn{1}{c|}{1.6}         & \multicolumn{1}{c|}{1.5}         & \multicolumn{1}{c|}{2.4}         & \multicolumn{1}{c|}{2.1}         & \multicolumn{1}{c|}{1.5}         & \multicolumn{1}{c|}{2.4}         & \multicolumn{1}{c|}{1.5}         & \multicolumn{1}{c|}{1.6}         & \multicolumn{1}{c|}{2.9}         & \multicolumn{1}{c|}{1.6}         & \multicolumn{1}{c|}{2.3}         & \multicolumn{1}{c|}{2.4}         & \multicolumn{1}{c|}{1.6}         & \multicolumn{1}{c|}{3.4}         & \multicolumn{1}{c|}{2.5}         \\ \hline
\multicolumn{1}{l}{}                      & \multicolumn{1}{l}{}             & \multicolumn{1}{l}{}             & \multicolumn{1}{l}{}             & \multicolumn{1}{l}{}             & \multicolumn{1}{l}{}             & \multicolumn{1}{l}{}             & \multicolumn{1}{l}{}             & \multicolumn{1}{l}{}             & \multicolumn{1}{l}{}             & \multicolumn{1}{l}{}             & \multicolumn{1}{l}{}             & \multicolumn{1}{l}{}             & \multicolumn{1}{l}{}             & \multicolumn{1}{l}{}             & \multicolumn{1}{l}{}             \\ \hline
\multicolumn{1}{|r|}{\textbf{Cluster}}    & \multicolumn{1}{c|}{\textbf{16}} & \multicolumn{1}{c|}{\textbf{17}} & \multicolumn{1}{c|}{\textbf{18}} & \multicolumn{1}{c|}{\textbf{19}} & \multicolumn{1}{c|}{\textbf{20}} & \multicolumn{1}{c|}{\textbf{21}} & \multicolumn{1}{c|}{\textbf{22}} & \multicolumn{1}{c|}{\textbf{23}} & \multicolumn{1}{c|}{\textbf{24}} & \multicolumn{1}{c|}{\textbf{25}} & \multicolumn{1}{c|}{\textbf{26}} & \multicolumn{1}{c|}{\textbf{27}} & \multicolumn{1}{c|}{\textbf{28}} & \multicolumn{1}{c|}{\textbf{29}} & \multicolumn{1}{c|}{\textbf{30}} \\ \hline
\multicolumn{1}{|r|}{\textbf{\#Images}}   & \multicolumn{1}{c|}{326}         & \multicolumn{1}{c|}{187}         & \multicolumn{1}{c|}{152}         & \multicolumn{1}{c|}{226}         & \multicolumn{1}{c|}{123}         & \multicolumn{1}{c|}{218}         & \multicolumn{1}{c|}{155}         & \multicolumn{1}{c|}{303}         & \multicolumn{1}{c|}{396}         & \multicolumn{1}{c|}{136}         & \multicolumn{1}{c|}{233}         & \multicolumn{1}{c|}{152}         & \multicolumn{1}{c|}{206}         & \multicolumn{1}{c|}{211}         & \multicolumn{1}{c|}{235}         \\ \hline
\multicolumn{1}{|r|}{\textbf{\#Patients}} & \multicolumn{1}{c|}{135}         & \multicolumn{1}{c|}{55}          & \multicolumn{1}{c|}{75}          & \multicolumn{1}{c|}{148}         & \multicolumn{1}{c|}{73}          & \multicolumn{1}{c|}{133}         & \multicolumn{1}{c|}{82}          & \multicolumn{1}{c|}{114}         & \multicolumn{1}{c|}{139}         & \multicolumn{1}{c|}{93}          & \multicolumn{1}{c|}{75}          & \multicolumn{1}{c|}{46}          & \multicolumn{1}{c|}{88}          & \multicolumn{1}{c|}{67}          & \multicolumn{1}{c|}{55}          \\ \hline
\multicolumn{1}{|r|}{\textbf{Ratio}}      & \multicolumn{1}{c|}{2.4}         & \multicolumn{1}{c|}{3.4}         & \multicolumn{1}{c|}{2.0}         & \multicolumn{1}{c|}{1.5}         & \multicolumn{1}{c|}{1.7}         & \multicolumn{1}{c|}{1.6}         & \multicolumn{1}{c|}{1.9}         & \multicolumn{1}{c|}{2.7}         & \multicolumn{1}{c|}{2.8}         & \multicolumn{1}{c|}{1.5}         & \multicolumn{1}{c|}{3.1}         & \multicolumn{1}{c|}{3.3}         & \multicolumn{1}{c|}{2.3}         & \multicolumn{1}{c|}{3.1}         & \multicolumn{1}{c|}{4.3}         \\ \hline
\end{tabular}
\end{table}
\end{appendix}

\end{document}